\documentclass[epj]{svjour}
\usepackage{latexsym}
\usepackage{graphics}
\begin{document}
\title{Entropy driven aggregation of adhesion sites of supported membranes}
\author{Noam Weil \and Oded Farago 
}                     
%
%
\institute{Department of Biomedical Engineering, Ben Gurion University,
Be'er Sheva 84105, Israel}
\date{Received: date / Revised version: date}
\abstract{
We study, by means of mean field calculations and Monte Carlo
simulations of a lattice-gas model, the distribution of adhesion sites
of a bilayer membrane and a supporting flat surface. Our model
accounts for the many-body character of the attractive interactions
between adhesion points induced by the membrane thermal
fluctuations. We show that while the fluctuation-mediated interactions
alone are not sufficient to allow the formation of aggregation
domains, they greatly reduce the strength of the direct interactions
required to facilitate cluster formation. Specifically, for adhesion
molecules interacting via a short range attractive potential, the
strength of the direct interactions required for aggregation is
reduced by about a factor of two to below the thermal energy $k_BT$.
%
     } 
%
\maketitle
\section{Introduction}
\label{intro}
Adhesion between two membranes or between a membrane and another
surface is an important topic for its ubiquitous occurrence in
biological and biophysical processes. This process, during which two
interfaces attract each other, can in principle be facilitated by
non-specific attractive interactions (e.g., Coulomb and van der Waals
interactions). Cell adhesion, however, is usually caused by highly
specific receptor molecules located at the outside of the plasma
membrane of the cell, that can bind to specific ligands on the
opposite surface \cite{beckerle:2001,lauff_inderman:1995}. Typically,
the area density of receptors is rather low which does not lead to
efficient bio-adhesion.  However, when facing a surface with enough
ligands, the receptors may cluster into highly concentrated adhesion
domains to establish much stronger binding
\cite{smith:2007,weikl:2009}. Formation of adhesion clusters occurs in
many biological processes \cite{lenne_nicolas:2009}, including the
binding of white blood cells to pathogens \cite{naggli:2003},
cadherin-mediated adhesion of neighboring cells
\cite{giehl_menke:2008}, and focal adhesion of cells to the
extracellular matrix \cite{geiger:2001}. Much insight into these
bio-adhesion processes has been gained from experimental studies of
biomimetic membranes with receptors molecules that interact with
surfaces covered by ligands
\cite{salafsky:1996,kloboucek:1999,cuvelier_nassoy:2004,tanaka_sackmann:2005,lamblet:2008,sengupta:2010}.

Generally speaking, adhesion induced domain formation requires some
attractive intermolecular interactions between the receptor-ligand
pairs. These interactions include both {\em direct}\/ and {\em
membrane-mediated}\/ contributions. The former are typically described
by pairwise interactions which are infinitely repulsive at very small
molecular separations and attractive at somewhat longer (but still
finite) distances \cite{israelachvili}. The effect of the direct
interactions between adhesion bonds can be studied in the framework of
the thoroughly researched lattice gas model
\cite{simon}. Specifically, there exists a critical value $\epsilon_c$
of the strength of the attractive part of the intermolecular pair
potential above which the system may phase separate into domains with
high and low concentrations of adhesion bonds.

Much less is known about the indirect interactions between adhesion
sites which are mediated by the membrane thermal fluctuations. The
entropic origin of these interactions can be easily understood as
follows: Consider two adhesion bonds between two membranes or between
a membrane and a surface. The adhesion points restrict the thermal
height fluctuations of the membrane in their vicinity. This entropy
loss can be minimized if the two adhesion bonds are brought to the
same place, in which case the membrane becomes pinned at only one
point and not two. The membrane fluctuations, thus, induce an
attractive potential of mean force between the adhesion points. 

In a previous publication, we analyzed the membrane mediated
interactions between two adhesion bonds of a bilayer membrane and a
supporting surface \cite{farago:2010}. We found that for
``point-like'' adhesion molecules whose size is comparable or smaller
than the thickness of the membrane ($l\sim 4-5$ nm), the potential of
mean force is an infinitely long range attractive potential that grows
logarithmically with the pair distance $r$:
\begin{equation}
U(r)=2k_BT\ln \left(\frac{r}{l} \right).
\label{eq:pairpotential}
\end{equation}
Eq.(\ref{eq:pairpotential}) holds for {\em tensionless}\/ membranes,
which will be the case discussed in the following. When surface
tension is applied, this form holds for pair separations
$r\ll\xi_{\sigma}=\sqrt{\kappa/\sigma}$, where $\kappa$ and $\sigma$
denote the membrane bending rigidity and surface tension,
respectively. For $r\gg\xi_{\sigma}$, $U(r)$ is decreased by a factor
of 2 \cite{farago:2010}. We leave the discussion in stressed membranes
to a future publication.

In this paper we consider the same system as in
ref.~\cite{farago:2010}, but instead of two adhesion points we look at
a membrane which is pinned to a flat impenetrable surface at multiple
sites. To analyze the aggregation behavior of the adhesion sites we
first need to generalize eq.(\ref{eq:pairpotential}) and write down
the fluctuation-induced interaction energy as a function of the
coordinates of the adhesion sites
$U(\vec{r}_1,\vec{r}_2,\vec{r}_3,\ldots,\vec{r}_N)$. This is a
non-trivial problem since the fluctuation-mediated interaction is a
many-body potential which cannot be expressed as the sum of two body
terms of the form in Eq.(\ref{eq:pairpotential}). The many-body nature
of $U(\vec{r}_1,\vec{r}_2,\vec{r}_3,\ldots,\vec{r}_N)$ is best
illustrated by an example: Consider a cluster of two adhesion points
located close to each other at $\vec{r}_1\simeq \vec{r}_2$, and a
third distant adhesion point located at $\vec{r}_3$. Having a single
adhesion point at $\vec{r}_1$ or $\vec{r}_2$ instead of the two-point
cluster will not result any change in the spectrum of membrane thermal
fluctuations. Therefore, the third point is attracted to the two-point
cluster by the same potential of mean force (\ref{eq:pairpotential})
to which it is attracted to one adhesion point, and not by a potential
which is twice larger than potential (\ref{eq:pairpotential}), which
would be the case if
$U(\vec{r}_1,\vec{r}_2,\vec{r}_3,\ldots,\vec{r}_N)$ is the sum of pair
interactions. What is the exact form of
$U(\vec{r}_1,\vec{r}_2,\vec{r}_3,\ldots,\vec{r}_N)$ is still an open
question that needs to be resolved for the fluctuation induced domain
formation to be understood. Several approximations to this problem
which avoid direct many-body calculations have been proposed. Weikl
and Lipowsky introduced a mean field theory in which the effect of the
pinning points is represented by a {\em homogeneous}\/ attractive
interaction between the fluctuating membrane and the underlying
surface (or between the membrane and another membrane)
\cite{weikl_lipowsky:2007}. They concluded that the homogeneous
fluctuation induced potential alone cannot facilitate the formation of
adhesion zones, but it greatly reduces $\epsilon_c$, the critical
strength of the direct interactions between the adhesion bonds above
which the formation of adhesion clusters is possible. A very similar
conclusion has been recently reached by Speck {\em et al.}\/ using
rigorous statistical mechanical methods \cite{speck:2010}. However, in
their model the hard wall interaction between the membrane and the
surface has been replaced by a harmonic confining potential.

In this work we present a more accurate approach to the problem which
employs a non-additive many body potential
$U(\vec{r}_1,\vec{r}_2,\vec{r}_3,\ldots,\vec{r}_N)$. The derivation of
the potential $U(\vec{r}_1,\vec{r}_2,\vec{r}_3,\ldots,\vec{r}_N)$,
which is based on our previous statistical mechanical studies of the
membrane thermal fluctuations with one \cite{farago:2008} and two
\cite{farago:2010} adhesion points, is presented in section
\ref{sec:manybody}. The idea is somewhat different than the one
previously used by others. As in
refs.~\cite{weikl_lipowsky:2007,speck:2010}, we integrate out the
membrane degrees of freedom and map the problem onto the lattice gas
model where the {\em occupied}\/ sites represent the adhesion
bonds. But unlike in refs.~\cite{weikl_lipowsky:2007,speck:2010}, the
interaction energy $U(\vec{r}_1,\vec{r}_2,\vec{r}_3,\ldots,\vec{r}_N)$
is not expressed as the sum of two body interactions between nearest
neighbor lattice sites. Instead, it is calculated from the {\em empty}
sites that represent the unpinned sections of the membrane. The energy
assigned with each empty site represents the loss of entropy due to
the reduction of the membrane fluctuations in the area around that
site. We argue that the extend by which the membrane fluctuations are
locally restricted depends mainly of the distance to the closest
pinning point. Therefore, the calculation of
$U(\vec{r}_1,\vec{r}_2,\vec{r}_3,\ldots,\vec{r}_N)$ for a given
distribution of pinning sites involves the division of the lattice
into Voronoi cells and summing their contributions to the free
energy. In section \ref{sec:meanfield} we present a mean field
analysis of our model, and in section \ref{sec:simulations} we present
the results of Monte Carlo lattice simulations. We find, in agreement
with other theoretical studies, that while the fluctuation-mediated
interactions alone are not sufficient to allow the formation of
aggregation clusters, they greatly reduce the strength of the residual
(direct) interactions between adhesion points which is required to
facilitate cluster formation. More specifically, we find that the
strength of the short range cohesive energy that allows condensation
is reduced by about a factor of two and falls below the thermal energy
$k_BT$. In section \ref{sec:summary} we summarize and discuss our
results.

\section{The many body fluctuation-mediated potential}
\label{sec:manybody}

\begin{figure}
\begin{center}
\resizebox{0.5\textwidth}{!}{%
  \includegraphics{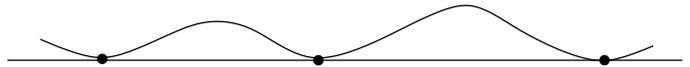}
}
\end{center}
\caption{Schematic of the system under investigation, consisting of a
membrane which is pinned to a flat impenetrable surface at multiple
adhesion points. The adhesion points can diffuse freely on the
surface, and they may cluster to increase the conformational entropy
of the fluctuating membrane.}
\label{figure_sketch}      
\end{figure}

We consider the system shown schematically in fig.~\ref{figure_sketch}
consisting of a fluctuating membrane of linear size $L$ which is
pinned to a flat impenetrable surface at several sites. The free
energy cost of a single adhesion point is given by
\cite{goulian:1994,farago:2008}
\begin{equation}
F_1=k_BT\ln\left(\frac{L^2}{l^2}\right)=2k_BT\ln\left(\frac{L}{l}\right).
\label{eq:singlepoint}
\end{equation} 
This result has been derived in ref.~\cite{farago:2008} by noting
that pinning the membrane to the surface at one point does not modify
the membrane spectrum of thermal fluctuations. It does, however,
eliminates the membrane translational degree of freedom by enforcing
the global minimum of the membrane height function to be located at
the point of contact with the surface. A different interpretation to
eq.(\ref{eq:singlepoint}) is that between the membrane and the surface
there is an effective interaction free energy per unit area of the
form
\begin{equation}
V(r)=\frac{k_BT}{\pi r^2},
\label{eq:singlepotential}
\end{equation}
that represents the entropy cost due to the reduction of the membrane
fluctuations at distance $r$ from the adhesion site
\cite{farago:2010}. The attachment free energy is then derived by
integrating the free energy density $U(r)$ over the projected area of
the membrane
\begin{equation}
F_1=\int_l^L\,2\pi rV(r)dr.
\label{eq:singlepoint2}
\end{equation}

The free energy density (\ref{eq:singlepotential}) can likewise be
employed to derive the fluctuation mediated pair potential
(\ref{eq:pairpotential}). However, when we have two adhesion points,
the distance $r$ in eq.(\ref{eq:singlepotential}) should be the
distance of the unit membrane area to the closest adhesion point. With
this choice of $r$, let us now consider a square membrane ($-L/2\leq
x,y\leq L/2$) with two adhesion points located at $(x,y)=(\pm
r_0/2,0)$. The attachment free energy is calculated by integrating
$V(r)$ over the projected area of the membrane (excluding a region of
size $l$ around the adhesion points):
\begin{eqnarray}
F_2&=&4\int_0^{L/2}dy\left[\int_0^{(r_0-l)/2}dx\,\frac{k_BT}
{\pi\left[y^2 +\left(x-r_0/2\right)^2\right]}\right.\nonumber \\
&+&\left.\int_{(r_0+l)/2}^{L/2}dx\,\frac{k_BT}{\pi\left[y^2
+\left(x-r_0/2\right)^2\right]}\right].
\label{eq:twopoints}
\end{eqnarray}
Integrating over $y$ yields, 
\begin{eqnarray}
F_2&=&\frac{4k_BT}{\pi}\left[\int_0^{(r_0-l)/2}\right. 
\frac{dx}{|x-r_0/2|}\tan^{-1}\left(\frac{L}{2|x-r_0/2|}\right)
\nonumber \\
&+&\int_{(r_0-l)/2}^{L/2}\left. 
\frac{dx}{|x-r_0/2|}\tan^{-1}\left(\frac{L}{2|x-r_0/2|}\right)
\right].
\label{eq:integrals}
\end{eqnarray}
Assuming that $l<r_0\ll L$, the inverse tangent function in
eq.(\ref{eq:integrals}) can be approximated by the constant value of
$\pi/2$ over most of the integration range (except near the boundaries
of the membrane $|x|\sim L/2$ which, nevertheless, does not influence
the dependence of the result on~$r_0$). With this approximation, one
gets
\begin{eqnarray}
F_2(r_0,L)&\simeq& 2k_BT\ln\left(\frac{L}{l}\right)
+2k_BT \ln\left(\frac{r_0}{l}\right)\nonumber \\
&=&F_1(L)+U(r_0).
\label{eq:twopoints2}
\end{eqnarray}
The first term in eq.(\ref{eq:twopoints2}) is the free energy cost of
a single adhesion site (\ref{eq:singlepoint}), which is the expected
value when the two adhesion points coincide ($r_0\simeq l$) to form a
single cluster. The second term, which represents the additional free
energy cost associated with the separation of the adhesion points, is
identified as the fluctuation induced pair potential, in agreement
with eq.(\ref{eq:pairpotential}).

We now wish to generalize eqs.(\ref{eq:twopoints}) and
(\ref{eq:twopoints2}), and propose that the many-body
fluctuation-mediated potential is given by
\begin{equation}
U\left(\vec{r}_1,\vec{r}_2,\vec{r}_3,\ldots,\vec{r}_N\right)=
F_N\left(\vec{r}_1,\vec{r}_2,\vec{r}_3,\ldots,\vec{r}_N,L\right)
-F_1(L).
\label{eq:manypotenial}
\end{equation}
The attachment free energy $F_N$ of $N$ adhesion points is calculated
by integrating the free energy density 
\begin{equation}
F_N=\int\frac{k_BT}{\pi 
d^2\left(\vec{r},\vec{r}_1,\vec{r}_2,\vec{r}_3,\ldots,\vec{r}_N\right)}
\,d\vec{r},
\label{npoint}
\end{equation}
where $d$ is the distance of the membrane unit area to the nearest
adhesion point
\begin{equation}
d=\min_{i=1\ldots N}\left(|\vec{r}-\vec{r}_i|\right),
\label{mindist}
\end{equation}
and the integration is carried over the projected area of the membrane
except for small regions of size $l$ near each adhesion point. What we
essentially argue here is that the extent by which the membrane
thermal fluctuations are limited at each spot depends almost
exclusively on the distance to the nearest adhesion point. All the
other adhesion points are effectively screened. This suggestion is
supported not only by the above example,
eqs.(\ref{eq:twopoints})-(\ref{eq:twopoints2}), but also by our
observation from recent computer simulations presented in
ref.~\cite{farago:2010}. In that paper, we determined the pair
potential (\ref{eq:pairpotential}) by using Monte Carlo simulations of
a coarse-grained bilayer model with two adhesion points. Our
simulations results were in excellent agreement with
eq.(\ref{eq:pairpotential}) despite of the fact that each adhesion
point interacts not only with the other adhesion point but also with
its infinite array of periodic images. It is unlikely that the
periodic images have such a negligible contribution on $U(r)$ for all
values of $r$ unless they are simply screened. In other words, the
local fluctuation behavior of the membrane is governed by the distance
to the nearest adhesion point while the spatial distribution of the
other, more distant, points is irrelevant. Thus, calculating $F_N$ for
a given distribution of adhesion points involves the construction of
the 2D Voronoi diagram of the configuration, and summation of the free
energy contributions coming from each Voronoi cell, $S_i$:
\begin{equation}
F_N=\sum_{i=1}^N\int dS_i\, \frac{k_BT}{\pi r^2},
\label{eq:voronoi}
\end{equation}
where $r$ is the distance to the adhesion point which is located
inside the Voronoi cell.

\section{Mean field theory}
\label{sec:meanfield}

To study the formation of adhesion clusters in supported membranes, we
consider a lattice of $N_s$ sites (with lattice spacing set equal to
$l$ (the cut-off microscopic length scale) of which $N<N_s$ sites are
occupied by adhesion points. For each spatial configuration of the
system, the energy is written as the sum of two terms
\begin{equation}
E=E_{\rm LG}+F_N.
\label{eq:latticeenergy}
\end{equation}
For the first term in eq.(\ref{eq:latticeenergy}) representing the
direct interaction between the adhesion points, we take the lattice
gas energy with nearest neighbor interactions
\begin{equation}
E_{\rm LG}=\sum_{\langle i,j\rangle} -\epsilon s_is_j,
\label{eq:latticegasenergy}
\end{equation}
where $s_i=1$ for an occupied site, $s_i=0$ for a vacant site, and
$\epsilon$ is the site-site interaction energy. The second term in
eq.(\ref{eq:latticeenergy}), which represents the attachment free
energy, is given by the discrete form of eq.(\ref{eq:voronoi})
\begin{equation}
F_N=\sum_{i=1}^{N_s}\frac{k_BT}{\pi}\left(\frac{l}{r}\right)^2(1-s_i),
\label{eq:latticevoronoi}
\end{equation}
where $r$ is the distance from a vacant lattice site to the nearest
occupied lattice site.

The phase behavior of the system can be analyzed through mean field
theory. Let us assume that the adhesion points form $N_c<N$
clusters. The free energy of system includes three contributions: (i)
the mixing entropy of the adhesion clusters, (ii) the energy $E_{\rm
LG}$ of the direct interactions between the adhesion points, and (iii)
the attachment free energy $F_N$. The first free energy contribution
is given by
\begin{equation}
\frac{F_{\rm
mix}}{k_BT}=N_c\left[\ln\left(\frac{N_c}{N_s}\right)-1\right]
+\frac{1}{2}c\left(\frac{N_c^2}{N_s}\right),
\end{equation}
where $c$ is the second virial coefficient. On average, each cluster
consists of $(N/N_c)$ adhesion points; and if we assume that it has a
roughly circular shape than $c\simeq 4(N/N_c)$. Denoting the number
densities of the adhesion points by $\phi=N/N_s$, and of the clusters
by $\phi^*=N_c/N_s\leq \phi$, the free energy of mixing per lattice
site is given by
\begin{equation}
\frac{F_{\rm mix}}{N_sk_BT}=\phi^*\left[\ln\left(\phi^*\right)-1\right]
+2\phi\phi^*.
\label{eq:free1}
\end{equation}

The second contribution to the free energy is due to the direct
interactions between the adhesion points.  The ground state of the
interaction energy $E_{\rm LG}$ is achieved when a single circular
adhesion domain with minimal surface is formed. If we set the ground
state as the reference energy, the energy of an ensemble of clusters
can be estimated as being proportional to the total length of the
domain boundaries. For $N_c$ circular clusters of size $(N/N_c)$ we
have
\begin{equation}
\frac{E_{\rm LG}}{N_sk_BT}=\lambda \frac{N_c}{N_s}\sqrt{\frac{N}{N_c}}
=\lambda\sqrt{\phi\phi^*},
\label{eq:free2}
\end{equation}
where $\lambda$, the associated dimensionless line tension, is
proportional to the interaction energy $\epsilon$ in
eq.(\ref{eq:latticegasenergy}) and $B$, the mean number of
nearest-neighbor vacant sites per occupied site on the boundary of a
cluster
\begin{equation}
\lambda=B\epsilon.
\label{b_parameter}
\end{equation}
The sum of free energy contributions (\ref{eq:free1}) and
(\ref{eq:free2}) constitutes the total free energy density (per lattice
site) of a 2D lattice gas of clusters:
\begin{equation}
\frac{F_{\rm LG}}{N_sk_BT}=\phi^*\ln(\phi^*)-\phi^*+2\phi\phi^*
+\lambda\sqrt{\phi\phi^*}.
\label{eq:freelg}
\end{equation}

The third contribution of the attachment free energy can be estimated
as follows. The clusters form $N_c$ Voronoi cells, each of which has on
average an area of $A_{\rm vor}=(N_s/N_c)l^2$. The attachment free
energy of each Voroni cell is given an equation similar to
eq.(\ref{eq:singlepoint}) for the attachment free energy of one
adhesion point, but with $A_{\rm vor}$ instead of the total membrane
area $L^2$. Thus
\begin{equation}
F_N=N_c\left[k_BT\ln\left(\frac{N_s}{N_c}\right)\right],
\end{equation}
and the attachment free energy density is given by
\begin{equation}
\frac{F_N}{N_sk_BT}=-\phi^*\ln(\phi^*),
\label{eq:free3}
\end{equation}
which eliminates the first term in the lattice-gas free energy density
[eq.(\ref{eq:freelg})], yielding
\begin{equation}
\frac{F}{N_sk_BT}=\frac{F_{\rm LG}}{N_sk_BT}+\frac{F_N}{N_sk_BT}=
-\phi^*+2\phi\phi^*
+\lambda\sqrt{\phi\phi^*}.
\label{eq:free}
\end{equation}

\begin{figure}[b]
\begin{center}
\resizebox{0.5\textwidth}{!}{%
  \includegraphics{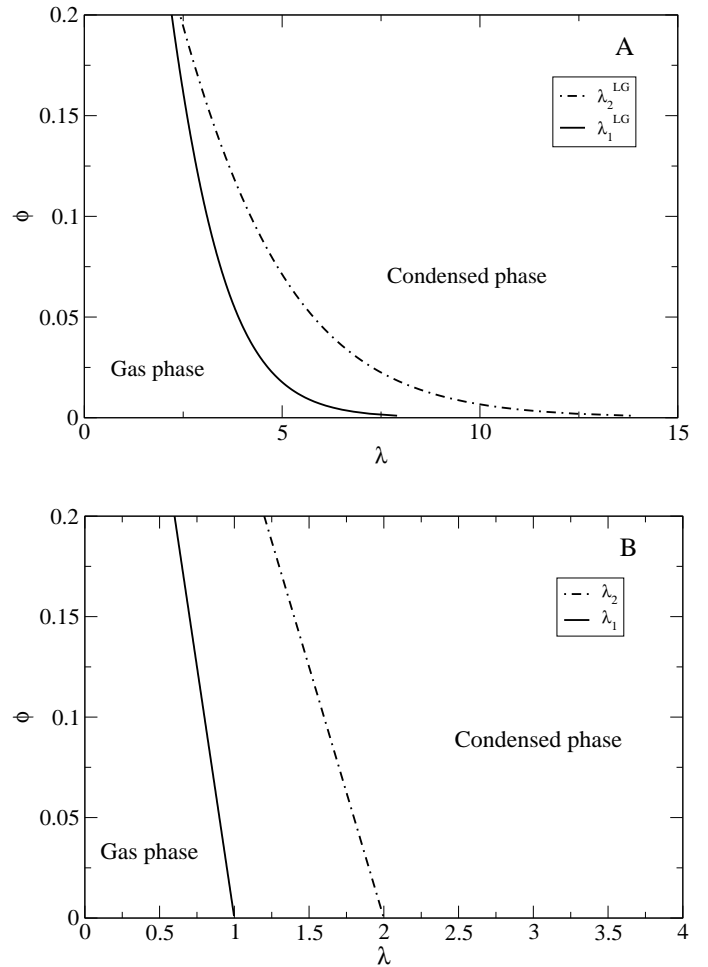}
}
\end{center}
\caption{The phase diagram of the adhesion sites calculated within the
mean field approximation. (A) {\protect Eq.(\ref{eq:lambdalg})} for
the lattice-gas model. (B) {\protect Eq.(\ref{eq:lambda})} for
adhesion points of fluctuating membranes. $\lambda_1$ and $\lambda_2$
represent the first-order transition and spinodal lines,
respectively.}
\label{figure_mean_field}      
\end{figure}

We consider a low density of adhesion sites $\phi\ll 1$, which also
implies a low number density of adhesion clusters since $\phi^*\leq
\phi$.  By minimizing the free energy density we obtain the equilibrium
value of the $\phi^*$ for the lattice-gas problem
[eq.(\ref{eq:freelg})] and for the adhesion points of a fluctuating
supported membrane [eq.(\ref{eq:free})]. In both cases, the system
undergoes a first order phase transition at $\lambda_1(\phi)$ from the
gas phase ($\phi^*=\phi$) to a condensed phase consisting of only a
few clusters ($\phi^*\sim 0$). Also, in both cases the free energy
reaches a maximum at intermediate densities ($0<\phi^*<\phi$). This free
energy barrier for condensation disappears at the spinodal point
$\lambda_2(\phi)>\lambda_1(\phi)$.  For the lattice-gas problem we
find
\begin{eqnarray}
\lambda_1^{\rm LG}&=& 1-2\phi-\ln(\phi) \nonumber \\ \lambda_2^{\rm
LG}&=& -4\phi-2\ln(\phi),
\label{eq:lambdalg}
\end{eqnarray}
while for the adhesion points of fluctuating membranes we have
\begin{eqnarray}
\lambda_1&=& 1-2\phi\nonumber \\ \lambda_2&=& 2-4\phi=2\lambda_1.
\label{eq:lambda}
\end{eqnarray}
The results of eqs.(\ref{eq:lambdalg}) and (\ref{eq:lambda}) are
summarized in fig.~\ref{figure_mean_field}A and B, respectively. The
important points in the results are that: (i) $\lambda_1>0$, which
means that the fluctuation induced interactions alone are not
sufficient to induce aggregation of adhesion domains, but (ii) they
greatly reduce the strength of the direct interactions required to
facilitate cluster formation since $\lambda_1<\lambda_1^{\rm LG}$ (and
also $\lambda_2<\lambda_2^{\rm LG}$). In the following section we
support these conclusions with MC simulations. We show that for
adhesion points of fluctuating membranes, the site-site cohesive
energy $\epsilon$ for the onset of aggregation falls below the thermal
energy $k_BT$.

\section{Computer simulations}
\label{sec:simulations}

\begin{figure}
\begin{center}
\resizebox{0.5\textwidth}{!}{%
  \includegraphics{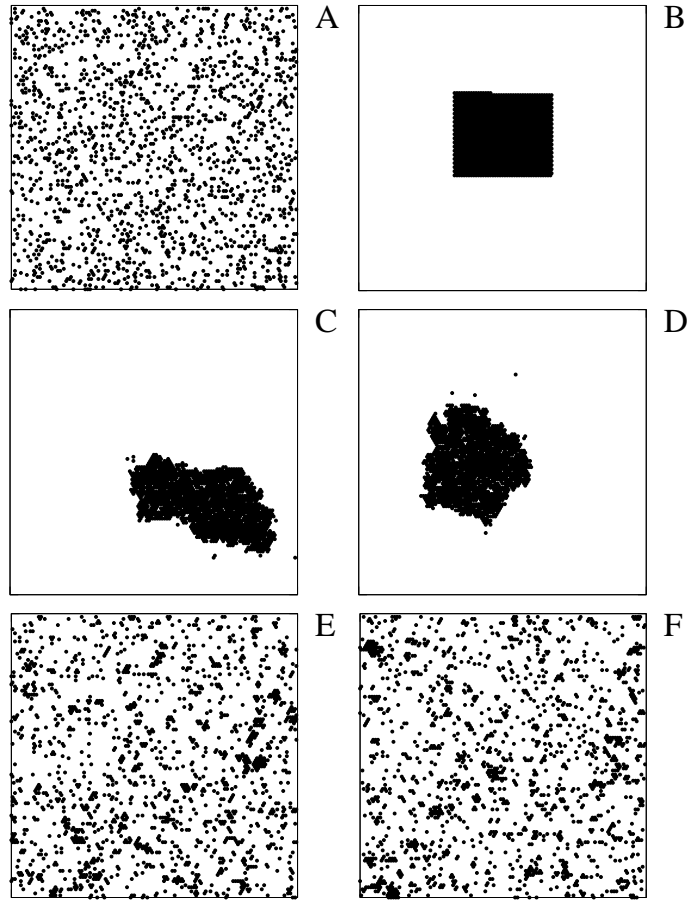}
}
\end{center}
\caption{Initial configurations of the simulations in which (A) the
sites are randomly distributed on the lattice, and (B) put in a single
compact cluster. Representative equilibrium configurations of (C-D)
our model [Eqs.(\ref{eq:latticegasenergy}) and
(\ref{eq:latticevoronoi})] and (E-F) the standard lattice-gas model
[Eq.(\ref{eq:latticegasenergy}) only] for $\phi=0.1$ and
$\epsilon=1k_BT$. Configurations (C) and (E) evolved from the initial
state (A), while (D) and (F) evolved from (B).}
\label{figure_snap}      
\end{figure}

To further investigate the aggregation behavior of adhesion points, we
performed Monte Carlo (MC) simulations of our lattice-gas model with
the total configurational energy given by the sum of direct
[eq.(\ref{eq:latticegasenergy})] and fluctuation-induced
[eq.(\ref{eq:latticevoronoi})] interactions. We used a $120\times138$
triangular lattice (that has an aspect ratio very close to 1) with
periodic boundary conditions. We simulated the system at two different
densities $\phi=N/N_s=0.05$ and $\phi=0.1$, and for various values of
$\epsilon$ ranging from 0 to 3 $k_BT$. For comparison, we also
simulated the standard lattice-gas model [for which the
configurational energy is given by eq.(\ref{eq:latticegasenergy}),
without the fluctuation-mediated free energy
eq.(\ref{eq:latticevoronoi})]. For each density $\phi$ and for each
value of $\epsilon$, we performed 8-16 independent runs starting from
different initial configurations where the points are either randomly
distributed on the lattice (as in fig.~\ref{figure_snap}A) or put in a
single cluster (see fig.~\ref{figure_snap}B). The system was then
equilibrated until the distribution of points in all the independent
runs look similar (see {\em e.g.,} fig.\ref{figure_snap}C {\em
vs.}~\ref{figure_snap}D, and \ref{figure_snap}E {\em
vs.}~\ref{figure_snap}F). Equilibrium time for the different samples
ranges from $3.6\times 10^5$ to $10^6$ MC time units, where each MC
time unit consists of $N$ single particle move attempts. For the
adhesion points problem, each particle was displaced to a randomly
chosen nearest neighbor lattice site, which enabled us to employ an
efficient algorithm to update the Voronoi diagram needed for
calculating the fluctuation-mediated free energy
(\ref{eq:latticevoronoi}). For the standard lattice-gas model, each
move attempt consisted of randomly selecting a particle and moving
it to the nearest vacant point in a randomly chosen lattice direction
\cite{remark}. After the first stage of equilibration, the simulations
were continued for $3\times 10^5$ MC time units during which data was
collected every third MC time unit.

\begin{figure}
\begin{center}
\resizebox{0.5\textwidth}{!}{%
  \includegraphics{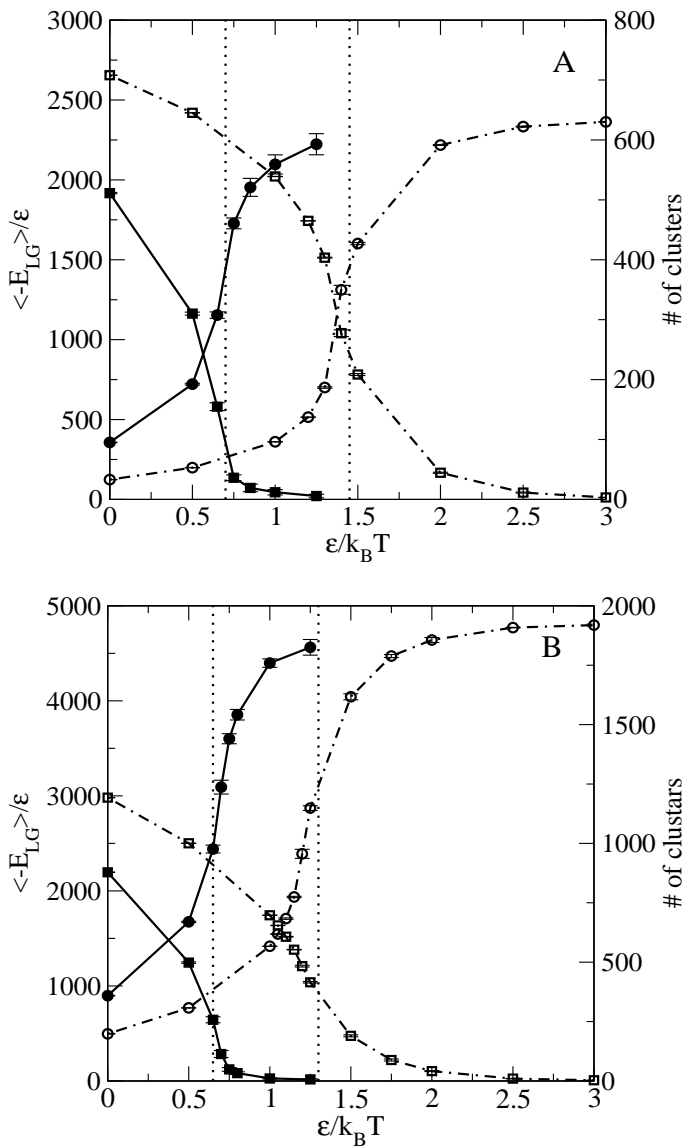}
}
\end{center}
\caption{Left $y$-axis: The energy of direct interactions between
sites, $\langle E_{\rm LG}\rangle$ {\protect
[Eq.(\ref{eq:latticegasenergy})]}, as a function of $\epsilon$, for
$\phi=0.05$ (A) and $\phi=0.1$ (B). Solid circles - results for our
model for adhesion points. Open circles - results for the standard
lattice-gas model. Right $y$-axis: The number of clusters as a
function of $\epsilon$, for $\phi=0.05$ (A) and $\phi=0.1$ (B). Solid
squares - results for our model for adhesion points. Open squares -
results for the standard lattice-gas model.}
\label{figure_sim}      
\end{figure}

To examine the occurrence of a phase transition from a gas to a
condensed phase, we measured the average number of clusters in the
system (where a cluster is defined as a set of neighboring occupied
sites), and the mean value of the energy of direct interactions
between sites, $\langle E_{\rm LG}\rangle$ [see
eq.(\ref{eq:latticegasenergy})]. Our results are summarized in
fig.~\ref{figure_sim}A (for $\phi=0.05$) and B (for $\phi=0.1$). For
each $\phi$ we measured these quantities both for the standard
lattice-gas model (open symbols and dash-dotted lines in
fig.~\ref{figure_sim}) and for adhesion points which also interact via
the fluctuation-mediated free energy eq.(\ref{eq:latticevoronoi})
(solid symbols and solid lines in fig.~\ref{figure_sim}). The number
of clusters is denoted by squares (values on the right $y$-axis of the
figures), while $\langle E_{\rm LG}\rangle$ is represented by circles
(values on the left $y$-axis).

The gas phase is characterized by a large number of small clusters,
some of which may be of the size of a single site. Furthermore, since
each occupied site has a relatively small number of neighboring
occupied sites, the mean configurational energy $\langle -E_{\rm
LG}\rangle$ is relatively low. Conversely, when the sites form large
clusters in the condensed phase, $\langle -E_{\rm LG}\rangle$ is high,
and the total number of clusters decreases (and in many cases,
especially for large values of $\epsilon$, we simply observe only a
single cluster in our system). Fig.~\ref{figure_sim} exhibits an
abrupt, clearly first-order, transition from a gas phase with a large
number of clusters and small $\langle -E_{\rm LG}\rangle$ to a
condensed state with a small number of clusters and large $\langle
-E_{\rm LG}\rangle$. The estimated values of $\epsilon$ at the
transition are (see vertical lines in fig.~\ref{figure_sim}):
$\epsilon_t\simeq0.7k_BT$ ($\phi=0.05$) and $\epsilon_t\simeq0.65k_BT$
($\phi=0.1$). In comparison (see also fig.~\ref{figure_sim}), for the
standard lattice-gas model, the transition values are roughly twice
larger than these values: $\epsilon_t^{\rm LG}\simeq1.45k_BT$
($\phi=0.05$) and $\epsilon_t^{\rm LG}\sim1.3k_BT$ ($\phi=0.1$).
Fig.~\ref{figure_snap}C-F exhibits typical equilibrium configurations
of the system at $\phi=0.1$ and $\epsilon=1k_BT$. For the lattice-gas
model $\epsilon_t^{\rm LG}>1 k_BT$, and at equilibrium the system is
in the gas phase (figs.~\ref{figure_snap}E and F). When the
fluctuation-induced interactions eq.(\ref{eq:latticevoronoi}) are
introduced, $\epsilon_t$ falls below $1k_BT$ and the system is in the
condensed phase where most of the particles belong to one large
cluster (figs.~\ref{figure_snap}C and D).

Our computational results which show that the fluctuation mediated
interactions reduce the strength of $\epsilon_t$, are in a qualitative
agreement with the mean field theory prediction (section
\ref{sec:meanfield}). To make a quantitative comparison between the
theory and the simulations, one needs to estimate the parameter $B$
appearing in eq.(\ref{b_parameter}). Several reasons make such an
estimation difficult and inaccurate: First, our non-standard mean
field theory is based on the assumption that the clusters are circular
and roughly have the same size, which is quite a crude
approximation. Second, tracing the precise location of $\epsilon_t$ in
fig.~\ref{figure_sim}) is largely inaccurate because of the finite
size of the system that makes the transition look like a crossover. To
reduce the large uncertainties associated with the determination of
$\epsilon_t$, one can look at the difference between the value of this
quantity in our model and for the standard lattice-gas model. Using
\begin{equation}
\lambda_1^{\rm LG}- \lambda_1=B\left(\epsilon_t^{\rm LG}-\epsilon_t\right),
\end{equation}
for $\phi=0.1$, we find $B\sim 3.5$ which for the triangular lattice
with 6 nearest-neighbors means that adhesion point residing on the
edge of a cluster loose slightly more than half of their neighbors.

\section{Summary and discussion}
\label{sec:summary}

In this paper we studied the aggregation behavior of adhesion points
between a fluctuating membrane and a supporting surface. We
demonstrated that the problem can be mapped onto a lattice-gas model
with two types of molecular interactions: (i) direct site-site pair
interactions and (ii) Casimir-like interactions which are mediated by
the membrane thermal fluctuations. The fluctuation-mediated
interactions, which are inherently of many-body character, are
calculated in our model by summing over the vacant rather than the
occupied sites of the lattice. Each vacant site represents a small
unit area of the fluctuating membrane, and the fluctuation-mediated
potential expresses the local free energy cost due to the restriction
imposed by the adhesion points on the membrane thermal
fluctuations. This free energy cost depends mainly on the distance
between the vacant sites and the nearest occupied site. Therefore,
such a many-body potential is calculated by determining the Voronoi
diagram for each lattice configuration, which can be quite easily
implemented in MC simulations.

We used mean field calculations and MC computer simulations to
investigate the phase behavior of a lattice-gas of adhesion sites at
low densities. We showed that upon increasing the strength of the
site-site interactions $\epsilon$, the system undergoes a first-order
phase transition into a condensed state. The fluctuation-induced
interactions lower the value of $\epsilon$ at the phase transition to
below the thermal energy $k_BT$. This result suggests that
fluctuation-mediated effects play a central role in the formation of
adhesion domains in biomimetic and biological membranes.

This work was supported by the Israel Science foundation (Grant Number
946/08).


\begin{thebibliography}{99}
\bibitem{remark} Such MC moves greatly reduce the rejection
probability due to excluded volume interaction. One can easily verify
that for these moves, detailed balanced is satisfied by the
conventional Metropolis acceptance rule {\protect $p({\rm
old}\rightarrow{\rm new})=\min(1,\exp(-\Delta E/k_BT))$}, where
$\Delta E$ is the energy change caused by the move attempt.
\end{thebibliography}
\end{document}